# A gold complex single crystal comprised of nanoporosity and curved surfaces

*Maria Koifman Khristosov*[(1-2)], *Leonid Bloch*[(1)], *Manfred Burghammer*[(3)], *Paul Zaslansky*[(4)], *Yaron Kauffmann*[(1)], *Alex Katsman*[(1)] *and Boaz Pokroy*[(1-2)]\*

(1)     Department of Materials Science and Engineering, Technion – Israel Institute of Technology, 32000 Haifa, Israel
(2)     The Russell Berrie Nanotechnology Institute, Technion – Israel Institute of Technology, 32000 Haifa, Israel
(3)     European Synchrotron Radiation Facility, BP 220, F-38043 Grenoble Cedex, France
(4)     Department for Restorative and Preventive Dentistry, Centrum für Zahn-, Mund- und Kieferheilkunde, Charité - Universitätsmedizin Berlin, 14197 Berlin

Complex hierarchical shapes are widely known in biogenic single crystals, but growing of intricate synthetic metal single crystals is still a challenge. Here we report on a simple method for growing intricately shaped single crystals of gold, each consisting of a micron-sized crystal surrounded by a nanoporous structure, while the two parts comprise a single crystal. This is achieved by annealing thin films of gold and germanium to solidify a eutectic composition melt at a hypoeutectic concentration (Au-enriched composition). Transmission electron microscopy and synchrotron submicron scanning diffractometry and imaging confirms that the whole structure was indeed a single crystal. A kinetic model showing how this intricate single-crystal structure can be grown is presented.

Prof. Boaz Pokroy
Department of Materials Science & Engineering
De-Jur building, Room 616
Technion - Israel Institute of Technology
Haifa 32000, Israel
Tel:  +972-4-8294584  (office)
Fax: +972-4-8295677
e-mail: bpokroy@tx.technion.ac.il
Homepage: http://pokroylab.net.technion.ac.il

# A gold complex single crystal comprised of nanoporosity and curved surfaces

*Maria Koifman Khristosov*[(1-2)], *Leonid Bloch*[(1)], *Manfred Burghammer*[(3)], *Paul Zaslansky*[(4)], *Yaron Kauffmann*[(1)], *Alex Katsman*[(1)] *and Boaz Pokroy*[(1-2)]\*

(1)  Department of Materials Science and Engineering, Technion – Israel Institute of Technology, 32000 Haifa, Israel

(2)  The Russell Berrie Nanotechnology Institute, Technion – Israel Institute of Technology, 32000 Haifa, Israel

(3)  European Synchrotron Radiation Facility, BP 220, F-38043 Grenoble Cedex, France

(4)  Department for Restorative and Preventive Dentistry, Centrum für Zahn-, Mund- und Kieferheilkunde, Charité - Universitätsmedizin Berlin, 14197 Berlin

\* bpokroy@tx.technion.ac.il

ABSTRACT

Complex hierarchical shapes are widely known in biogenic single crystals, but growing of intricate synthetic metal single crystals is still a challenge. Here we report on a simple method for growing intricately shaped single crystals of gold, each consisting of a micron-sized crystal surrounded by a nanoporous structure, while the two parts comprise a single crystal. This is achieved by annealing thin films of gold and germanium to solidify a eutectic composition melt at a hypoeutectic concentration (Au-enriched composition). Transmission electron microscopy and synchrotron submicron scanning diffractometry and imaging confirms that the whole structure was indeed a single crystal. A kinetic model showing how this intricate single-crystal structure can be grown is presented.

**INTRODUCTION**

Biogenic crystals are often formed via amorphous precursors rather than via classical nucleation and growth.[1-3] With this approach, various organisms are able to grow unfaceted sculptured single crystals with complex morphologies such as rounded shapes, highly porous structures, and hierarchical intricate forms. In contrast, crystals grown synthetically by classical methods of nucleation and growth exhibit facets dictated by atomic structure and minimization of surface energy.[4] This results in faceted crystals in which the revealed planes are the low-energy ones.

There are two approaches to the production of synthetic single crystals with intricate shapes – a "bottom-up" approach and a "top-down" approach. Using the "bottom-up" approach, a single crystal can be grown in a template of the desired shape, for example by growing ceramic single crystals in polymer templates.[5, 6] This approach is widely used in ceramic materials due to the fact that the growth of the crystals is performed from a solution. In the "top-down" approach single crystals are produced by etching or mechanical treatment, such as in dealloying[7] or polishing.[8] In the case of gold, commonly used processes to create nanoporous gold are by dealloying from a gold-silver alloy[9, 10] or by liquid metal dealloying.[11-13] However, these methods enable to create nanoporosive bulk samples or thin layers of gold, however without other possible morphologies and shapes. Polishing can be used for macro-manipulation but is ineffective for creating nanoporosity. Thus, creation of elaborate and combined structures of single crystals of metals remains a challenge.

Here we describe a method for growing micron-sized particles comprised of gold single crystals with intricate morphologies, consisting partially of a whole single crystal, and partially of a nanoporous single crystal. For this purpose we exploited the phase diagram of

gold−germanium, which is eutectic and has a relatively low eutectic temperature ($T_E$ = 361 °C).[14] Previously we showed that nanoporous single crystals of gold can be grown by using the eutectic concentration.[15] We further found that by using a hypoeutectic concentration (Au-enriched composition) we could grow curved single crystals of gold.[16] In the present work however, we show that by using the eutectic gold−germanium system at hypoeutectic concentrations it is possible to grow single-crystal gold particles with intricate morphologies consisting partially of full and partially of nanoporous gold.

For this purpose, gold and germanium thin films are evaporated on a nonreactive surface at thicknesses yielding a hypoeutectic concentration in the phase diagram. Heating above the liquidus line of the eutectic phase diagram leads to melting and dewetting of gold and germanium, with isolated droplets of the melt produced on the substrate's surface. Growth of the intricate single-crystal structure of gold in these droplets occurs in two stages: (i) growth of a single crystal in the melt droplet (above the eutectic temperature), and (ii) crystallization of the eutectic melt into a eutectic microstructure (at the eutectic temperature or below). The process is illustrated in detail in Figure 1.

Thus, the first step in growing single-crystal gold particles with complex morphologies is achieved by cooling the sample and crossing the liquidus line to undergo the phase transition L → L + $Au_{(s)}$. In each droplet, excess gold is crystallized into a single crystal. At this point each droplet consists of a single crystal surrounded by eutectic liquid.

In the second step the sample is cooled below the eutectic temperature to undergo the phase transition L + $Au_{(s)}$ → eutectic $AuGe_{(s)}$ + $Au_{(s)}$. The remaining eutectic liquid in each droplet around the single crystal solidifies into a eutectic microstructure. The gold lamellas in the eutectic microstructure grow homo-epitaxially on the surface of the single crystal. In this way the gold lamellas become a continuation of the former single crystal, and a complex single-crystal structure is formed.

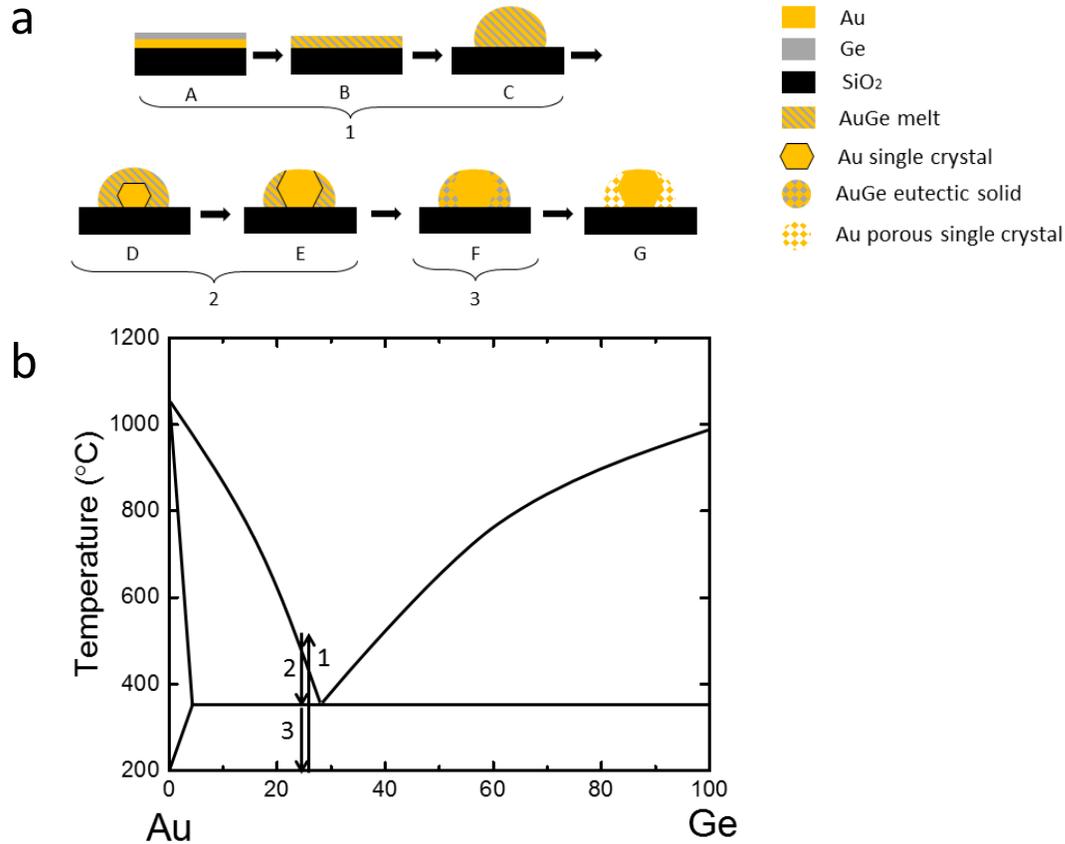

**Figure 1.** (a) Schematic illustration of the gold micro-particle production. A. Thin film evaporation of gold and germanium on a Si/SiO₂ substrate. B. Annealing above the liquidus line and melt formation. C. Melt droplets formation due to dewetting. D. Cooling from the liquidus line to the eutectic temperature – gold single crystal is growing inside the droplet. E. The single crystal continues to grow. F. Cooling from the eutectic temperature to room temperature – the melt around the single crystal crystallizes into a eutectic microstructure. G. Selective etching of germanium. (b) Au-Ge phase diagram, the process in (a) is marked with arrows. Arrow 1 corresponds to stages A, B and C, arrow 2 to stages D and E, and arrow 3 is F. Selective etching of Ge from the particles reveals a complicated single-crystal structure of gold consisting of both a whole single crystal and a nanoporous single crystal.

We developed a model that explains our experimental findings. The model shows that the diffusion-controlled epitaxial growth of gold on the large crystal occurs more rapidly than heterogeneous crystallization with multi-nucleation points, which would result in whole crystal and nanoporous gold with different orientations.

**EXPERIMENTAL SECTION**

*Sample Preparation*. Silicon dioxide (100 nm) was grown on (001) Si wafers by thermal oxidation at 1100 °C. The oxide layer served as a diffusion barrier preventing migration of Si from the wafer. Gold and germanium films (99.999% pure, Sigma-Aldrich) were successively evaporated onto the $SiO_2$ substrate in an e-beam-equipped AircoTemescal FC-1800 evaporating system under a high vacuum of $10^{-7}$ Torr at room temperature, yielding a deposition rate of 8 Å$s^{-1}$. The gold and germanium films were 65 nm and 26 nm thick, respectively. Samples were thermally annealed in forming gas ($N_2 + H_2$) at 650 °C for 10 min in a Linkam TS-1000 Hi-Temp Stage. After the latter step the samples were wet-etched by immersion in a solution of $NH_4OH:H_2O_2$ (1:25vol%) for 45 min and finally rinsed in deionized water. Cross-sectional samples were obtained by an FEI Strata 400S dual-beam FIB. Low-voltage argon ion milling was then applied for final thinning and cleaning of the surface using the Gentle Mill, model IV8 (Technoorg Linda). NaCl single crystals that were used as substrates were one side optically polished and were purchased from CRYSTAL GmbH.

*Sample Characterization*. Surfaces were imaged with a Zeiss Ultra Plus HRSEM. Cross sections were micrographed with the FEI Strata 400S dual-beam FIB. We performed TEM imaging and obtained electron diffractions using an FEI Titan 80−300 KeV Cs-corrector FEG-S/TEM operated at 300 KeV. Single-crystalline droplets were characterized by nanofocus X-ray beam analysis on ID13 of the European Synchrotron Radiation Facility (ESRF), Grenoble, France. The sample was measured at a single angle, using an X-ray beam focused to approximately 200×150 nm at FWHN. The wavelength (λ) used was 0.832053 Å. The number of frames was 121×51, the number of pixels per frame was 1024×1024, and the 2×2 binning was used. The exposure time was 1s.

The 3D morphology of intact samples was obtained by imaging an array of particles on the ID19 imaging beamline of the ESRF. Sub-micron CT scans were obtained at an energy of 26 KeV with the particles mounted in a PVA vial, positioned as close to the detector as physically possible (~10 mm). The optical configuration resulted in an effective pixel size of 0.16 μm. Reconstruction was performed using conventional filter back projection using PyHST2.[14]

**RESULTS AND DISCUSSION**

Thin films of gold and germanium, 65 nm and 26 nm thick, respectively (23at% Ge, hypoeutectic concentration), were evaporated on a silicon dioxide surface. The choice of $SiO_2$ as a substrate enabled droplets of this melt to be formed by dewetting. Annealing was performed at 650 °C for 10 min in forming gas. Dewetting during the annealing process led to the formation of Au−Ge droplets under which single gold crystals were formed. Annealing was followed by rapid cooling of the samples to room temperature at the rate of 200 °Cmin$^{−1}$, which led to solidification of the eutectic melt around the single gold crystal in each droplet. During this latter process the eutectic melt solidified into a eutectic microstructure around the gold crystal. As a result, solid droplets with a combined microstructure were formed on the surface (Figure 2).

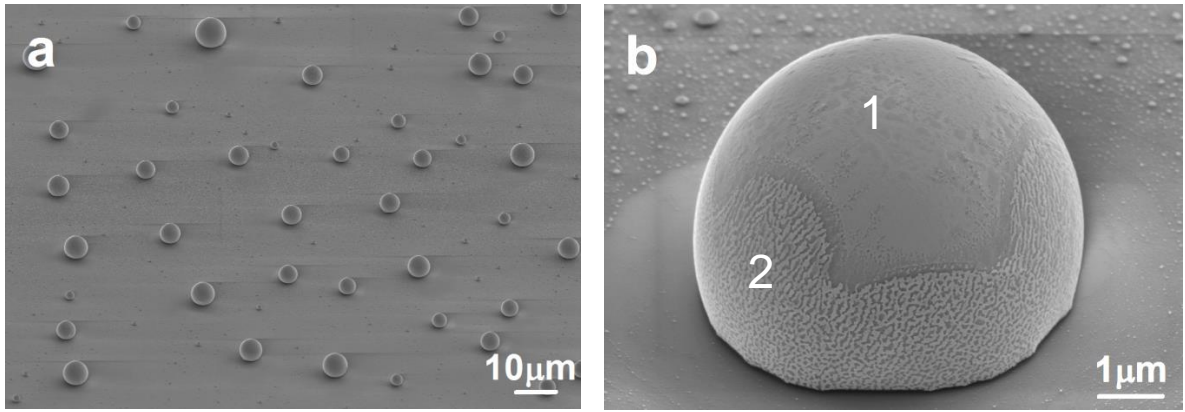

**Figure 2.** High-resolution scanning electron microscopy (HRSEM) images of Au−Ge droplets after annealing. (a) Large area view (52° tilt). (b) Side view (60° tilt). Mark 1 is for the gold crystal, mark 2 is for the eutectic microstructure of gold-germanium.

At this point we wanted to investigate the crystallinity and the orientation of the nanoporous structure relative to the large crystal. We prepared a thin cross section of a sample droplet using a focused ion beam (FIB) and examined it by transmission electron microscopy (TEM). A high-angle annular dark-field scanning transmission electron microscopy (HAADF-STEM) image of the cross section is presented in Figure 3a.

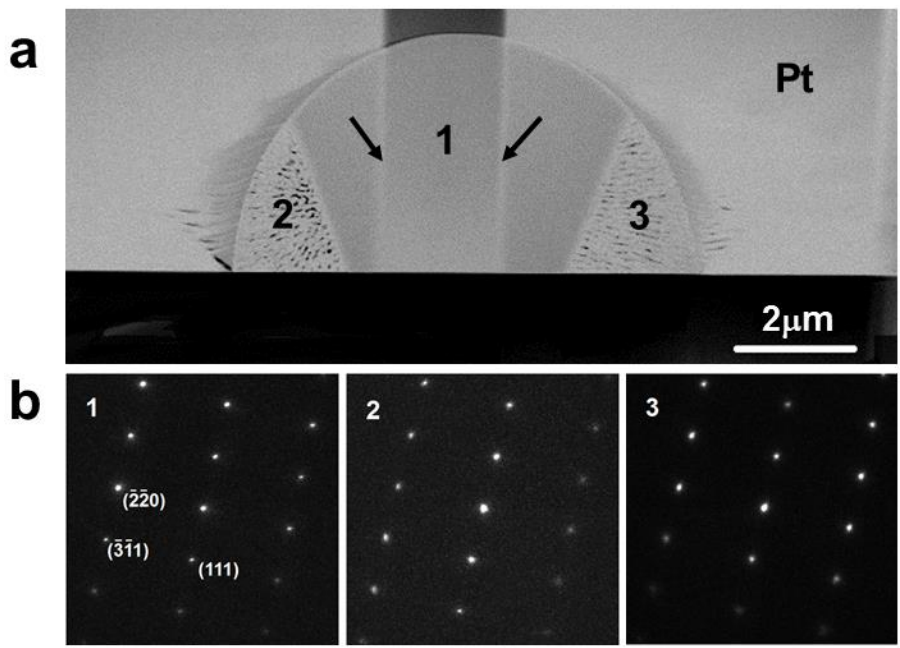

**Figure 3.** (a) A high-angle annular dark-field HAADF-STEM image of the droplet's cross section, covered with platinum for protection during sample preparation by FIB. Area 1 contains a whole gold single crystal. Areas 2 and 3 are porous single crystals of gold, oriented as in area 1. The lines (marked with small arrows) are caused by milling of the single-crystal area to thinner lamellae for better diffraction of the single crystal (area 1). (b) Diffractions taken from areas 1−3 in (a), with similar orientations, could be fully indexed within the gold structure (zone axis [121]).

Selected area diffraction in different parts of the droplet from a large area (0.5 μm diameter) yielded a diffraction pattern of single-crystalline nature that could be fully indexed within the gold structure. We obtained identical electron-diffraction patterns from different locations of the droplet's cross section (Figure 3b). The diffraction patterns were taken from the dense gold single crystal in the center (marked as 1) and from the gold porous regions on both the left and the right sides of the single crystal (marked as 2 and 3). From these findings we concluded that all three areas have the same crystallographic orientation, indicating that the gold in the eutectic-like structure was solidified homo-epitaxially onto the pure single crystal as a continuation of the micron-sized single crystal in the center of the droplet.

Additional proof that all the different parts of the droplet have the same orientation was obtained by means of another state-of-the-art characterization technique, namely synchrotron-based submicron scanning diffractometry (ID13, European Synchrotron Radiation Facility (ESRF), Grenoble, France), on an FIB-sectioned gold crystal.[17] We used this technique to examine a similar cross section to that of the droplet investigated by TEM. The entire cross section of the droplet was scanned using a sub-micron X-ray beam and diffraction patterns were collected from the various points. In this manner 2844 scans were obtained from the droplet area.

Average diffraction images of the different zones can be seen in Figure 4a−c. Figure 4a depicts the average of all the diffractions taken from the entire area of the droplet. This can be compared with Figure 4b, which shows an average diffraction of the porous zone only, and with Figure 4c, which shows an average diffraction of the whole single crystal. By comparing these images, we can see that the major reflections are identical and do not shift their positions even though the figure integrates a large number of individual scans. Stronger background can be seen in Figure 4c (relatively to the porous areas in the crystal) due to stronger fluorescence in the area of the whole single crystal. This is due to higher amount of gold on the path of the X-rays.

A map of a selected single reflection {220} in all zones can be seen in Figure 4d, and maps of the same {220} reflection of each zone separately can be seen in Figure S1. All squares that contain a diffraction spot are equal in size, and are taken from exactly the same radial and azimuthal coordinates of a diffraction image, the origin of which on the drop corresponds to the location of the square on the map. From these mappings it is clear that the intensity is maintained at roughly the same level for the {220} reflection, and that the spots are located at the same reciprocal coordinates throughout the drop, both in the porous and the solid areas.

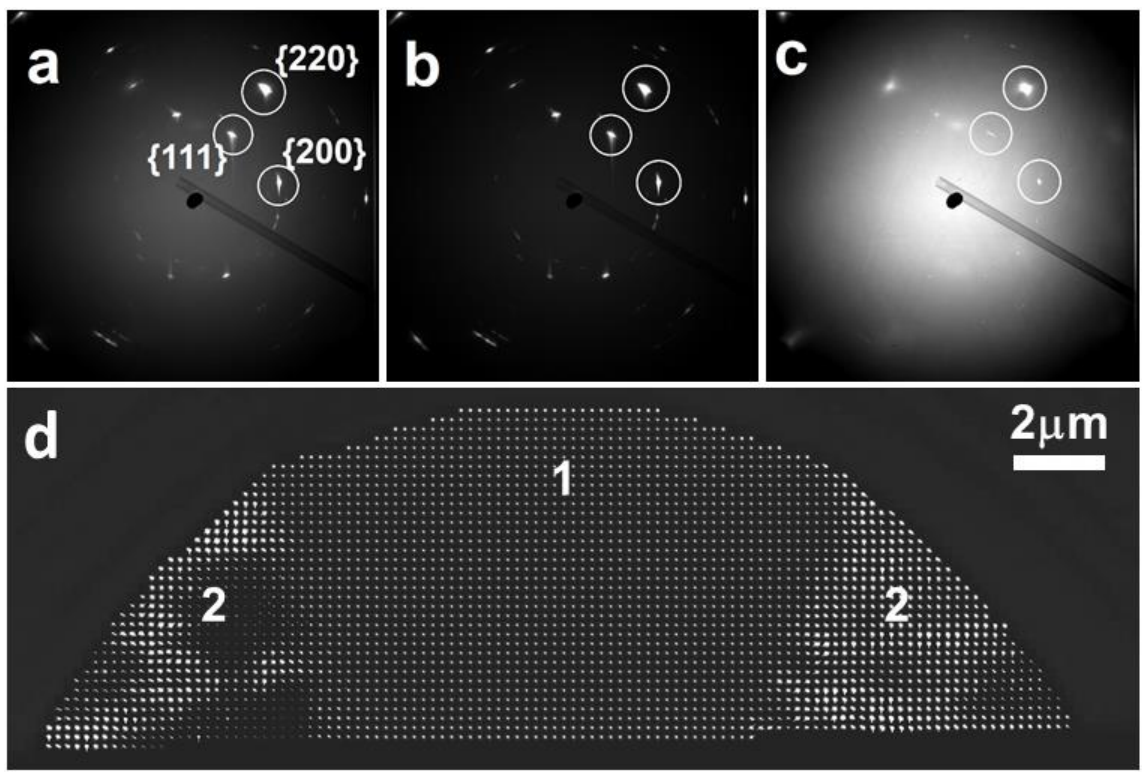

**Figure 4.** (a) Average diffraction of the entire drop (both regions). Selected reflections are marked. (b) Average diffraction of the porous zone only. (c) Average diffraction of the central zone only. (d) Map of a single reflection {220} from the entire drop. Area 1 is the central zone of the whole single crystal, area 2 is the porous zone.

The slight inhomogeneity which can be seen in the porous areas can be explained by the damage caused by FIB preparation, during which the gentle regions of the sample might have been slightly bent.

The results obtained from TEM and X-ray scanning diffractometry provide conclusive proof that the intricately structured gold single-crystal particles made partially of full and partially of nanoporous gold are indeed single crystals.

An additional characterization we performed on these crystals was sub-micron synchrotron X-ray µCT (ID19, ESRF). For this purpose, the same type of gold particles was grown on a NaCl single crystal substrates to facilitate easy detachment by simply immersing the samples in water. Collected particles were then imaged by HRSEM as can be observed in Figure 5a. X-ray sub-micron CT of such particles revealing the faceted single crystal within a porous coat are shown in Figure 5b-d and Movie S1. The full crystal can clearly be seen within the droplet, exhibiting a shape of a truncated octahedron.

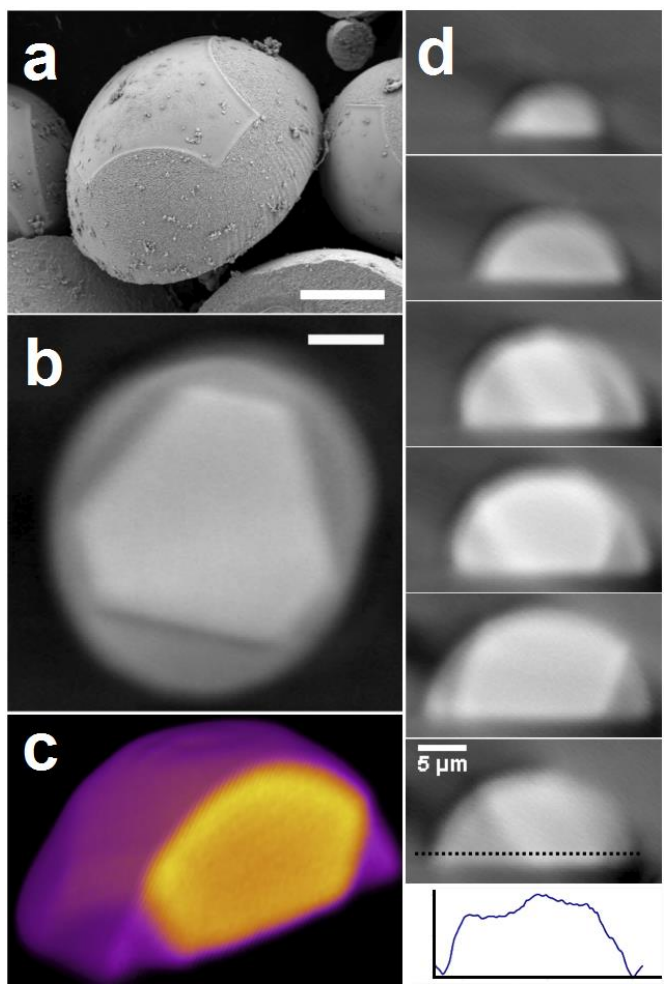

**Figure 5**. All scale bars are 5 µm. (a) Detached, intricately structured gold single-crystal particles, HRSEM image. (b) X-ray sub-micron CT of a typical gold single crystalline particle. An enlarged top view cross-section of the similar particle as in (a), observed 2.5 µm above the base. The faceted core completely encircled by the nanoporous gold area. (c) An example 3D rendering of the full tomographic reconstruction of one of the droplets. False colors highlight the different densities. (d) A series of slices along the particle reveals the cross-sectional relations between the high-density gold region and the surrounding less-dense nanoporous crystal. Profiles across such slices (lower right panel, marked with a dashed line on the droplet) reveal that the density of the surrounding material is 20~40% lower than the density of the central crystal. All images in this section have the same magnification and are presented with the same intensity scale, clearly revealing the higher density core.

Our TEM and synchrotron results show that the outcome of the described growth process is an intricately shaped single crystal of gold, consisting of a micron-sized crystal surrounded by a nanoporous structure, while the two parts comprise a single crystal. The processes and kinetics involved can be explained as follows.

The crystallization of a gold crystal in the melt droplet happens according to the phase diagram. It is important to understand how the eutectic melt crystallizes around the initially formed gold single crystal. Crystallization of Au-Ge eutectics in the presence of a solid gold single crystal may start and progress in different ways. [i] One way is through nucleation of a two-phase Au/Ge unit comprising a new gold crystal (with random orientation with respect to the former gold single crystal) and an adjacent germanium crystal at the solid/liquid interface, followed by their simultaneous growth into the eutectic liquid. [ii] Another way is through local diffusion-controlled growth of the existing gold single crystal from the eutectic liquid accompanied by substantial enrichment of the adjacent liquid by germanium, providing heterogeneous nucleation of Ge crystal at the surface of the Au crystal.

To evaluate these possibilities, we need to compare the time needed in the two cases for full crystallization of the eutectic structure. This process includes nucleation events and subsequent eutectic growth. The time needed for eutectic growth over the whole droplet is the same in both cases, so we actually need to compare only the times needed for nucleation of the eutectic units.

We start by evaluating the time required for heterogeneous nucleation of a two-phase Au/Ge unit with a gold crystal misoriented in relation to an existing single gold crystal (case [i], Figure S2). The nucleation rate can be deduced on the basis of classical kinetic theory:[14, 18]

$$J_{ss} = J_0 \exp\left(-W^* / kT\right) \tag{1}$$

where $J_0$ is a pre-exponential factor and $W^*$ is the height of the heterogeneous nucleation barrier. Evaluation of $W^*$ for the two-phase Au-Ge cubic nucleus is presented in Note S1.

The time $t_1$ required for heterogeneous nucleation of the first nucleus can be determined by integration of the nucleation rate (Equation (1)) with time:

$$\Delta V_d \int_0^{t_1} J_0 \exp\left(-\frac{W^*}{k_B T}\right) dt = 1 \qquad (2)$$

where $\Delta V_d$ is the near-surface volume appropriate for heterogeneous nucleation. Since the nucleation occurs over a narrow temperature interval[19] we can assume $T \approx T_e$ ($T_e$ is the eutectic temperature) is constant in the exponent's denominator, while the undercooling $\Delta T$ has to be considered as the time-dependent parameter $\Delta T = T_e - T = \Delta T_0 + \alpha t$, where $\Delta T_0$ is the undercooling formed during preliminary formation of the main gold single crystal and $\alpha$ is the cooling rate. Integration of Equation (2) is presented in Note S2, and therefore the time required for nucleation of the first two-phase cubic-like nucleus is found to be $t_1 = (2.9 \div 5.8)\,\text{s}$. Therefore, substantial undercooling ($\Delta T = \alpha t_1 \sim (100 \div 200)\,°C$) is required for a first nucleation of such a two-phase particle. It should be noted that undercooling of this large magnitude is not usually reached in this system.

We now calculate the time required for eutectic unit nucleation in case [ii], where a new gold layer grows diffusively on the surface of an existing gold crystal with the same crystal orientation, followed by nucleation and growth of the Ge layer (Figure S3). In this case there is no energy barrier to nucleation of a new gold crystal, while a Ge crystal should heterogeneously nucleate from the Ge-rich liquid near the solid gold.

Growth of a gold crystal from the eutectic liquid in the reaction $L(X_e) \rightarrow L'(X_{L'}) + S(Au)$, below the eutectic temperature $\Delta T = (T_{eut} - T)$, is governed by the free energy change:

$$\Delta g_{Au} = \frac{X_{L'} - X_e}{X_{L'}} \Delta g_f^{Au} - T \left[ \frac{X_e}{X_{L'}} S_L^{conf}(X_{L'}) - S_L^{conf}(X_e) \right] + \frac{X_e}{X_{L'}} g^E(X_{L'}) - g^E(X_e) \qquad (3)$$

where $X_e$ and $X_L$ are the molar fractions of Ge in the eutectic liquid and in the liquid L', $\Delta g_f^{Au}$ is the free energy of fusion of Au; $S_L^{conf}(X_e)$ and $S_L^{conf}(X_{L'})$ are the configuration entropies in the liquid before (at $X_e$) and after gold crystallization (at $X_{L'}$); and $g^E$ is the excess free energy in the liquid. The first term in Equation (3) is responsible for nucleation of a gold crystal and the other two terms for supersaturation of the melt with Ge, which then provides nucleation of a Ge crystal. Using the thermodynamic modeling,[14] the free energy gain $\Delta g_{Au}$ can be approximated as follows (see Note S3):

$$\Delta g_{Au} \approx -\frac{\Delta T}{m_{Au} X_e}(a + b\Delta T) \qquad (4)$$

where $a = 255.8$ Jmol$^{-1}$, $b = 4.56$ Jmol$^{-1}$·K$^{-1}$, and $m_{Au} = 22$ K/at.% is the slope of the Au liquidus line (defined as positive) in the Au−Ge phase diagram. Diffusional growth of a new gold layer on the surface of the existing gold crystal results in the energy change:

$$\Delta g = \Delta g_{Au} \lambda^2 h + 4\gamma_{Au-L} \lambda h \qquad (5)$$

where $h$ is the layer thickness. The process will take place when $\Delta g$ is negative, and this will occur if the new layer size is larger than a critical value:

$$\lambda > \lambda_{min}^{Au} = -\frac{4\gamma_{Au-L}}{\Delta g_{Au}} = \frac{4\gamma_{Au-L} m_{Au} X_e}{\Delta T(a + b\Delta T)} \qquad (6)$$

For the parameters determined above, the value $\lambda_{min}^{Au} \sim 80$ nm is reached for undercoolings $\Delta T = (40 \div 60)$K (Figure S4) and this value is comparable to our results.

Advancement of the new gold layer of width λ inside the eutectic liquid is accompanied by an enrichment of the melt with Ge, especially near the Au/melt interface. Nucleation of Ge crystals from the melt is driven by the difference in Ge chemical potentials:

$$\Delta g_{Ge} = \mu_{Ge}^{L}(X_L) - \mu_{Ge}^{S} \qquad (7)$$

Using thermodynamic modeling,[14] we obtain:

$$\Delta g_{Ge} = \Delta g_f^{Ge} + R_g T \ln X_L + (1-X_L)^2 \left[ L_0 + L_1(1-4X_L) + L_2(1-2X_L)(1-6X_L) \right] \qquad (8)$$

where $\Delta g_f^{Ge}$ is the free energy of fusion of Ge and $R_g$ is the gas constant.

Amplitudes of concentration fluctuations in front of the solid/liquid interface moving at the rate $V$ can be evaluated as $\Delta X_0 = \lambda V / D$, where $D$ is the coefficient of Au diffusion in the undercooled eutectic liquid ($\Delta X_o$ corresponds to the coefficient $B_1$ in Ref. [20]). For reasonable values of λ=0.1 μm, $V$=40 μms$^{-1}$ and $D$=40 μm$^2$s$^{-1}$ we can evaluate $\Delta X_o \approx 0.1$. The nucleation driving force $\Delta g_{Ge}$ (Equation (7)) as a function of $\Delta X_o = X_L - X_e$ is presented in Figure S4. The ratio $\varphi = \Delta g_{Ge} / \Delta g_f^{Ge}$ can be seen to reach the value 0.2÷0.4 for supersaturations 0.05÷0.11. The time needed for heterogeneous nucleation of the first Ge crystal nucleus can be estimated as follows:

$$t_2 = \left( J_0 \Delta V_d \right)^{-1} \exp\left[ \frac{\kappa \tilde{\Gamma}^3}{2 k_B T \varphi^2 \left( T_m^{Ge} - T \right)^2 \left( \Delta s_f^{Ge} \right)^2} \right] \qquad (9)$$

where $\tilde{\Gamma} = 2(2\gamma_{Ge-L} + \gamma_{Au-Ge} - \gamma_{Au-L})$ and $\Delta s_f^{Ge}$ = 3 MJm$^{-3}$·K$^{-1}$ is the fusion entropy of Ge. Using typical values for $J_0 \approx (4\div6) \cdot 10^{20}$ s$^{-1}$ μm$^{-3}$, $\Delta V_d \approx a \cdot \left( \lambda_{min}^{Au} \right)^2 \approx 2 \cdot 10^{-6}$ μm$^3$, and $\tilde{\Gamma} \approx 0.7$ Jm$^{-2}$, for $T$ = 600 K we can obtain $t_2 < 10^{-4}$ s for $(X_L - X_e) > 0.05$ ($\varphi > 0.2$) (Figure S5).

Bearing these estimates in mind, we can conclude that nucleation of germanium crystals at the surface of the new gold layer occurs rapidly and does not retard the eutectic growth.

The diffusion-controlled growth of a germanium layer adjacent to a new gold layer results in an energy decrease, if its size exceeds a certain critical value $\lambda > \lambda_{min}^{Ge}$, as calculated in Note S4:

$$\lambda^{Ge} > \lambda_{min}^{Ge} = \frac{2(2\gamma_{Ge-L} + \gamma_{Au-Ge} - \gamma_{Au-L})m_{Ge}X_e}{\Delta T(a' + b'\Delta T)} \tag{10}$$

The even smaller minimum crystal size in the case of Ge (compared to Au) (Figure S4) can be explained by a larger energy gain in the subsequent crystallization of Ge from the melt enriched by Ge after primary crystallization of gold. As can be seen from the calculations, the diffusion-controlled growth (case [ii]) of the eutectic structure with the typical spacing can be obtained for undercooling above 50 K, if the gold part of the eutectics is a continuation of the existing gold single crystal.

The eutectics growth rate in the Au–Ge system has been evaluated based on the theory of Jackson and Hunt[20-22] as $V = 0.295/\lambda^2$ μms$^{-1}$=30÷50 μms$^{-1}$.[15] Therefore, full crystallization of the eutectic structure over the whole micron-sized droplet occurs within 0.001÷0.01 s at a near-constant temperature, when undercooling reaches a certain critical value corresponding to an operating point, with spacing $\lambda = \phi\lambda_{min}$, where $\lambda_{min} = \lambda_{min}^{Au} + \lambda_{min}^{Ge}$ and $\phi > 1$ is a constant reflecting the spacing adjustment mechanism.[20] The near-constant temperature of the entire eutectics crystallization results in a very weak dependence of the eutectic structure on the cooling rate in the investigated range 1÷35 °Cs$^{-1}$.[15] In the next step, gold is epitaxially crystallized from the Au-enriched liquid, and the eutectic crystallization process continues.

Thus, the nucleation of a two-phase Au/Ge eutectic unit, with the Au orientation possibly differing from that of the former gold single crystal (case [i]), requires 3÷6 s for the first nucleation event, while the local diffusion-controlled growth of the existing gold single crystal followed by Ge crystallization (case [ii]) requires very short times of <0.01 s. Bearing in mind the high rates of the eutectic growth, we can conclude that the diffusion-controlled growth of the eutectic structure, with the gold part being the continuation of the existing single gold crystal, occurs and completes the crystallization of the whole droplet within a very short period (a few hundredths of a second) at near-constant undercooling $\Delta T = 40 \div 60$ K, long before the possible nucleation of a new Au/Ge unit. That is why the large gold crystal and the porous gold structure in the droplet are, in fact, the same single crystal.

**CONCLUSION**

To conclude, we showed in this work that intricate structures of gold composed of single crystals can be grown by the relatively simple method of exploiting crystallization from a hypoeutectic melt. By using TEM and synchrotron submicron scanning diffractometry and imaging we confirmed that the whole structure is indeed a single crystal. We presented a kinetic model that shows how this intricately-structured single crystal can be grown.

**ASSOCIATED CONTENT**

**Supporting information**

The supporting information is available free of charge via the Internet at http://pubs.acs.org.

Additional calculations for the discussion. Maps of a selected single reflection {220} of different zones of the single crystals. Schematic diagram of nucleation of a two-phase plate-like particle at the surface of an existing gold single crystal. Schematic diagram of the diffusion-controlled growth of a new gold crystal from the eutectic Au−Ge melt at the surface of an existing gold single crystal. Minimum sizes of gold and germanium crystals providing a barrier-free growth as a function of undercooling. The energy gain due to crystallization of Ge from undercooled supersaturated eutectic liquid and the time needed for heterogeneous nucleation of the Ge crystal nucleus as a function of liquid supersaturation $(X_L - X_e)$ for undercooling $\Delta T = (T_{eut} - T) = 34$ K.

## AUTHOR INFORMATION

**Corresponding Author**

*E-mail: bpokroy@technion.ac.il

**Author Contributions**

The manuscript was written through contributions of all authors. All authors have given approval to the final version of the manuscript.

**Notes**

The authors declare no competing financial interest

## ACKNOWLEDGEMENTS

Thin films were fabricated at the Micro-Nano Fabrication Unit (MNFU) at the Technion—Israel Institute of Technology, Haifa. We thank Dr. Tzipi Cohen-Hyams, Dr. Alex Berner, and Michael Kalina for their help in preparing samples and operating the electron microscopes. The research leading to these results was awarded funding from the European Research Council under the European Union's Seventh Framework Program (FP/2007–2013)/ERC Grant Agreement (no. 336077). M.K.K. acknowledges with thanks the financial support from the Israeli Ministry of Science, Technology and Space. Nanofocusing lenses were kindly provided by the group of Prof. Dr. Christian Schroer from the Technische Universität Dresden, Germany. The diffraction and imaging experiments were performed on beamlines ID13 and ID19 at the European Synchrotron Radiation Facility (ESRF), Grenoble, France. We are grateful to Dr. Alexander Rack at the ESRF for providing assistance in using beamline ID19.

A gold complex single crystal comprised of nanoporosity and curved surfaces

Maria Koifman Khristosov, Leonid Bloch, Manfred Burghammer, Paul Zaslansky, Yaron Kauffmann, Alex Katsman and Boaz Pokroy

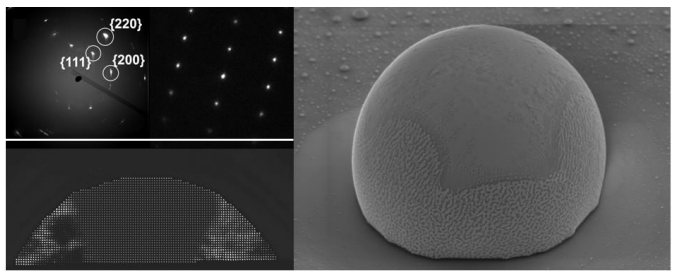

or

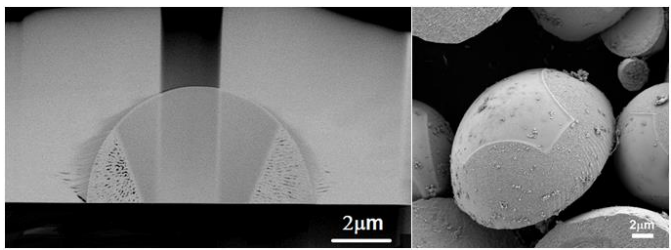

Or

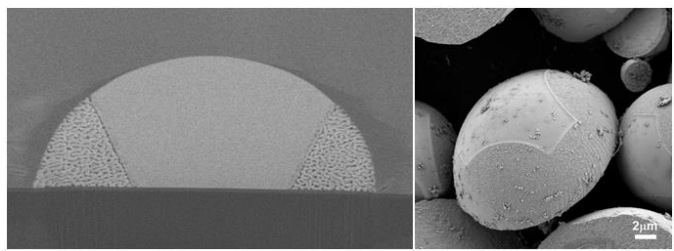

**Synopsis**

Intricately shaped gold single crystals, each consisting of a micron-sized crystal surrounded by a nanoporous structure, were grown. These two parts comprise a single crystal. This was achieved by growing gold crystal from a hypoeutectic melt of a gold-germanium eutectic system. TEM and synchrotron submicron scanning diffractometry and imaging confirmed that the whole structure was indeed a single crystal.